\documentclass[sigconf, preprint]{acmart}
\AtBeginDocument{%
  }

\usepackage{subcaption}
\renewcommand\footnotetextcopyrightpermission[1]{} % removes footnote with conference information in first column
\makeatletter
\renewcommand\@formatdoi[1]{\ignorespaces}
\def\@eprintstripzeros#1{}
\makeatother

\acmConference[Working Paper]{}{July 2026}{Toronto, Canada}

\begin{document}

% \title[Learning with Multi-Agent LLM Configurations]{Beyond One-on-One: Multi-Agent LLM Configurations for Convergent and Divergent Learning}
\title[Human Thinking under Plural LLM Assistance]{Human Thinking under Plural LLM Assistance: Mathematical Problem Solving and Open-Ended Writing}

\author{Harsh Kumar}
\email{harsh@cs.toronto.edu}
\orcid{0000-0003-2878-3986}
\affiliation{%
  \institution{University of Toronto}
  \city{Toronto}
  \state{Ontario}
  \country{Canada}
}

\author{Zi Kang (Jace) Mu}
\email{jace.mu@mail.utoronto.ca}
\orcid{0009-0000-4933-5416}
\affiliation{%
  \institution{University of Toronto}
  \city{Toronto}
  \state{Ontario}
  \country{Canada}
}

\author{Jonathan Vincentius}
\email{jon.vincentius@mail.utoronto.ca}
\orcid{0009-0007-2443-1576}
\affiliation{%
  \institution{University of Toronto}
  \city{Toronto}
  \state{Ontario}
  \country{Canada}
}

\author{Ashton Anderson}
\email{ashton@cs.toronto.edu}
\orcid{0000-0003-3089-6883}
\affiliation{%
  \institution{University of Toronto}
  \city{Toronto}
  \state{Ontario}
  \country{Canada}
}

\renewcommand{\shortauthors}{Kumar et al.}

\begin{abstract}
Large language models are changing not only the kind of assistance people receive, but also how that assistance is organized. Instead of working with a single general-purpose chatbot, people can now receive help from systems arranged as peers, specialists, or multiple agents with distinct roles. However, it remains unclear how these forms of plural LLM assistance affect human performance, confidence, and diversity of thought. We conducted two controlled experiments involving 562 participants to examine the effects of using multiple LLMs on mathematical problem-solving and writing. In a math task, participants worked with no LLM, an expert assistant, peer-like agents that surfaced common errors, or both an expert and a peer-like assistant. The expert-plus-peer condition produced the strongest unassisted post-task performance. In a writing task, participants wrote with no LLM, a single generalist assistant, or a pair of role-specialized assistants. LLM assistance improved essay quality, but the role-specialized pair preserved greater idea diversity than the single assistant. Together, these findings identify the arrangement of LLM assistance as a consequential design variable for human-AI collaboration.
\end{abstract}




\maketitle

\section{Introduction}

Large language models are becoming part of everyday thinking \cite{huang2026interviewer, chatterji2025people}. People now turn to them for help solving problems, learning concepts, drafting arguments, and generating ideas. As a result, a growing body of work has begun to study how LLM assistance affects human cognition: whether it improves performance, changes what people learn, alters their confidence, or shapes the ideas they produce \cite{lee2025impact, kumar2024human, tankelevitch2024metacognitive}.

Much of this work has naturally focused on the interaction between one person and one assistant. This has been a useful starting point. A single chatbot is easy to study, easy to deploy, and closely matches how many people first encounter LLMs. But it captures only one way that AI assistance can be organized. LLMs can be instantiated as one voice or many, as generalists or specialists, as tutors, peers, critics, collaborators, or panels \cite{lu2026assistant}. The same model can be prompted to play different roles or personas \cite{shanahan2023role}, different models can be placed side by side, and multiple agents can agree, disagree, or divide a task among themselves \cite{llm_council, estornell2024multi}. This is becoming increasingly practical as people compare answers across models, ask different systems for different kinds of help, and use interfaces that combine multiple model outputs \cite{Thomas_2026}.

On one hand, plural assistance could expose users to alternative strategies, common mistakes, complementary perspectives, or role-specific feedback. Such arrangements may help people reason more carefully, feel more confident, or generate a broader range of ideas than they would with a single assistant. On the other hand, more help may not necessarily be better help. Additional agents may increase cognitive load, fragment attention, or encourage users to offload more of the task. We therefore ask how the use of multiple LLMs affects human thinking across two kinds of tasks: mathematical problem solving, where people work toward a correct answer, and open-ended writing, where both quality and diversity matter. Specifically, we ask:

\begin{description}
    \item[RQ1:] In math problem solving, how do expert-like, peer-like, and combined arrangements of LLM assistance affect confidence and later unassisted performance?
    \item[RQ2:] In open-ended writing, how do single-generalist and role-specialized plural arrangements of LLM assistance affect confidence, essay quality, and idea diversity across writers?
\end{description}

To answer these questions, we conduct two controlled experiments. In the first, participants solve mathematical problems with no AI, an expert assistant, peer-like agents that surface common errors, or both an expert and a peer-like assistant. Participants first work through supported lesson problems and then solve related problems without AI assistance, allowing us to examine whether different arrangements affect later independent performance rather than only in-the-moment support. In the second experiment, participants write open-ended essays with no AI, a single generalist assistant, or a pair of role-specialized assistants. This allows us to examine whether plural assistance changes not only writing quality and confidence, but also the diversity of ideas that different writers produce.

Across these experiments, we find that plural assistance can matter, but not simply because more AI is always better. In the math task, the expert-plus-peer arrangement produced the strongest unassisted post-task performance. In the writing task, AI assistance improved essay quality, but the role-specialized pair preserved greater idea diversity than the single assistant. Together, these findings suggest that the cognitive consequences of LLM assistance depend not only on model capability or response quality, but also on the arrangement of assistance around the user.

\section{Related Work}

\subsection{Single Assistants to Plural LLM Arrangements}
A growing body of work studies how LLM assistance affects human thinking, including performance, learning, confidence, cognitive effort, metacognition, trust, reliance, and creativity \cite{lee2025impact, tankelevitch2024metacognitive, kumar2024human, bo2024disclosures, kumar2024guiding, kim2024m, kim2025fostering}. Much of this work studies a single user interacting with a single assistant. However, this captures only one point in a broader design space. LLM assistance can vary not only in model capability or response quality, but also in how support is arranged, e.g., one voice or many, one model or several, generalist or specialist roles, stable assistant personas, and agents that agree, disagree, or divide labor \cite{shanahan2023role, lu2026assistant, wu2024autogen, li2023camel, park2023generative, li2023metaagents}.

This broader design space has important precedents. Research in collaborative learning, peer learning, and computer-supported collaborative learning has long examined how learners benefit from explanation, peer modeling, shared problem solving, and exposure to others' reasoning \cite{dillenbourg1999you, webb1989peer, roscoe2008tutor, chi2008observing, slavin1980cooperative, johnson2003social, roschelle1995construction, schunk1987peer}. Prior AI systems have also explored interaction structures beyond one-on-one tutoring, including virtual peers, teachable agents, facilitator agents for collaborative learning, and configurable conversational agents \cite{cassell2000shared, leelawong2008designing, matsuda2010learning, chaudhuri2009engaging, kumar2010architecture, tegos2011mentorchat}.

Our work builds on this tradition. LLMs make it easier to instantiate multiple agents, assign roles or personas, vary model identity, and deploy these arrangements across everyday cognitive tasks. At the same time, LLM roles and personas should be treated as interfaces and prompting choices rather than as human-like social actors \cite{shanahan2023role, cheng2024one, devrio2025taxonomy}. We therefore study plural LLM assistance as a configurable arrangement of support around the user, asking whether this arrangement affects performance, confidence, and diversity of thought.

\subsection{LLM Assistance for Mathematical Problem Solving}
Math problem-solving is a useful setting for studying AI assistance because success depends not only on completing the current problem, but also on what people can do after assistance is removed. Intelligent tutoring systems and conversational tutors have long shown that hints, feedback, explanation, and adaptive support can improve learning in structured domains \cite{bloom19842, carbonell1970ai, anderson1995cognitive, koedinger1997intelligent, vanlehn2011relative, nye2014autotutor, d2013autotutor}. Recent LLM-based work extends this line of research, but the evidence suggests that the effects of generative AI depend strongly on how assistance is provided. LLM-generated explanations and feedback can support performance and later problem-solving in controlled learning platforms \cite{kumar2023math, pardos2024chatgpt, lira2025learning}, while open-ended or poorly scaffolded access to generative AI can encourage answer-seeking, reduce cognitive effort, or harm later learning \cite{bastani2024generative, lee2025impact, tankelevitch2024metacognitive, yan2025beyond}. 

Several mechanisms suggest why arrangement may matter in mathematical problem solving. Prior work shows that learners can benefit from explaining, observing others' reasoning, diagnosing errors, and encountering manageable confusion or productive struggle \cite{roscoe2008tutor, chi2008observing, siegler2006microgenetic, vanlehn1999rule, kapur2008productive, kapur2010productive, d2012dynamics}. These mechanisms motivate peer-like agents that surface common arithmetic or conceptual errors. However, they do not imply that more assistance is always better: additional sources of support can increase cognitive load, introduce conflicting information, or make it easier for learners to offload reasoning \cite{bjork1994memory, bastani2024generative, lee2025impact}. We therefore compare no AI, an expert-like assistant, peer-like agents, and a combined expert-plus-peer arrangement, measuring confidence and unassisted post-task performance.

\subsection{Homogenization in LLM Writing}
Open-ended writing raises a different question. Here, AI assistance may improve the quality of individual outputs while reducing variation across people. Homogenization here refers to reduced diversity across users' outputs. Essays may become more similar in their claims, themes, examples, narrative elements, or phrasing. Prior work has found that language model assistance can reduce content diversity in writing \cite{padmakumar2023does}, and related work finds that AI-generated ideas can improve evaluations of individual creative outputs while changing the diversity of ideas produced across a population \cite{anderson2024homogenization, ashkinaze2025ai, kumar2025human}. More broadly, research on human-AI writing and group creativity shows that external suggestions, shared prompts, and group interaction can shape both individual quality and collective diversity \cite{lee2022coauthor, paulus2003enhancing, paulus2003group, nijstad2006group, diehl1987productivity}.

A single generalist assistant may pull many writers toward similar ideas, whereas multiple assistants or role-specialized assistance may expose writers to a wider range of possibilities. At the same time, additional agents can increase distraction or cognitive load. Our writing experiment, therefore, compares no AI, a single generalist assistant, and a pair of role-specialized assistants, measuring writing quality, confidence, and idea-level diversity across writers.

\begin{figure*}[t]
  \centering
  \includegraphics[width=\textwidth]{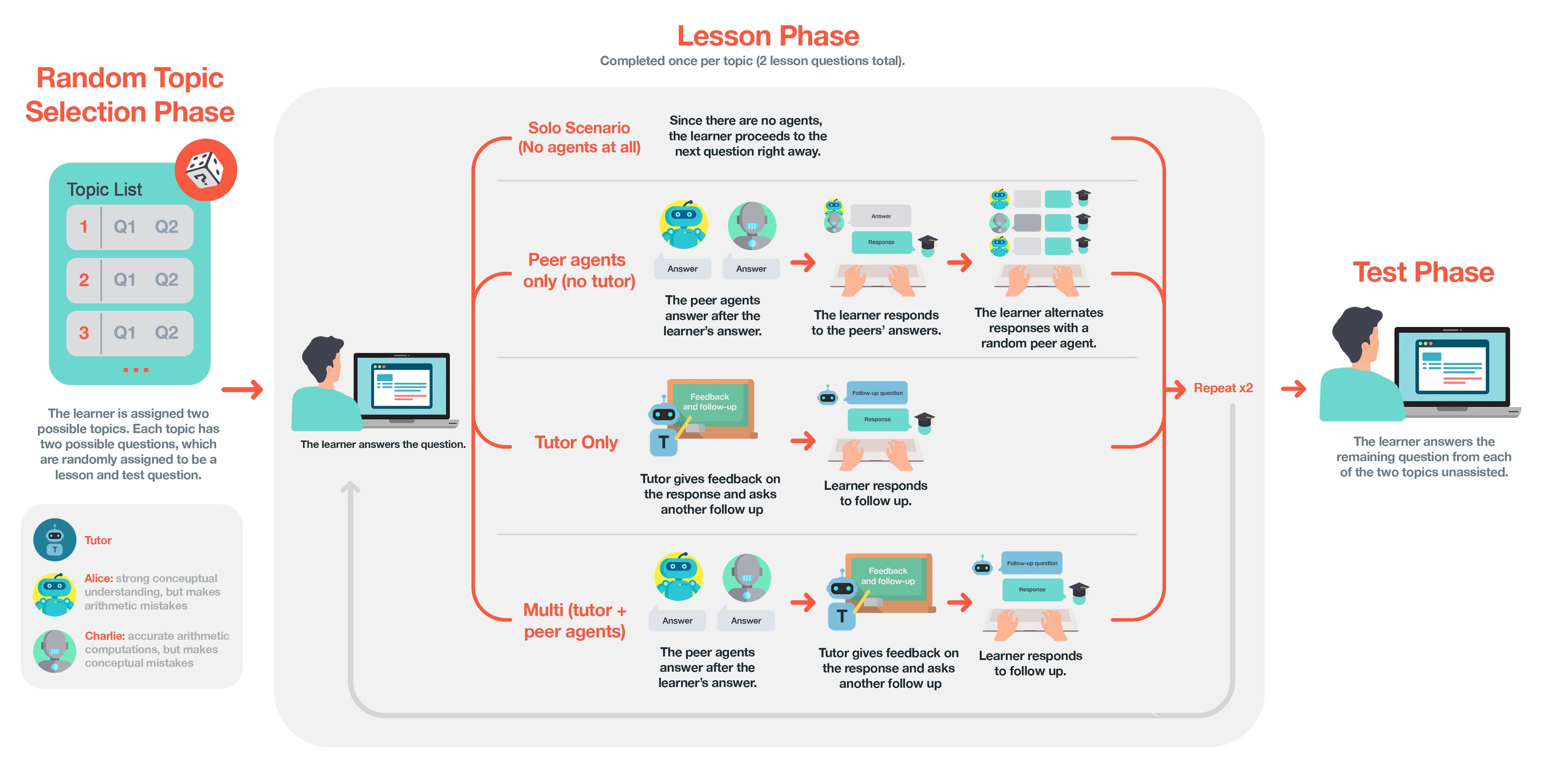}
  \caption{Experimental procedure and conditions for Experiment-1. Participants first go through a random topic-selection step that assigns two topics, each with two possible questions (one designated for the lesson phase and one for the test phase). In the lesson phase, participants answer one question per topic (two lesson questions total) under one of four support conditions: \emph{Control} (no agents), \emph{Peers only} (two peer agents answer after the participant), \emph{Tutor only} (a tutor provides feedback and a follow-up question), or \emph{Tutor + Peers} (peers respond after participant's answer, then the tutor provides feedback and a follow-up). The lesson sequence repeats twice (once per topic). In the test phase, participants answer the remaining question from each topic unassisted (two test questions total).}
  \Description{Diagram of the study flow. Left: a random topic-selection phase assigns two topics, each with two candidate questions. Center: a lesson phase in which the participant answers one question per topic with varying support: (1) Control with no agents; (2) Peers only, where two peer agents answer after the participant and the participant responds; (3) Tutor only, where a tutor gives feedback and asks a follow-up question that the participant answers; and (4) Tutor plus peers, combining peers’ answers with subsequent tutor feedback and a follow-up. The lesson phase repeats twice, once per topic. Right: a test phase where the participant answers the remaining questions from each topic without any support.}
  \label{fig:exp-learning-procedure}
\end{figure*}

\section{Experiment-1: Math Problem-Solving with LLM Tutor and Peers}
We designed a pre-registered\footnote{\url{https://aspredicted.org/rq8uz9.pdf}} controlled experiment to investigate how different configurations/arrangements of LLM agents may affect learners' math problem-solving (Figure \ref{fig:exp-learning-procedure}). The focus here was on a structured problem-solving domain in which retention and self-confidence were key measures (important components of human thinking).

\subsection{Experimental Design}
\subsubsection{Design of LLM Agents}

All agents were based on GPT-5.2 with distinct system prompts (provided in the Appendix). We designed three agents, each playing a different role in the learning interaction:

\texttt{Bob (Tutor).} Bob acts as a supportive math tutor who guides the discussion and provides feedback on the participant's answers. Rather than supplying solutions directly, Bob responds with clarifications, hints, and encouragement (consistent with Socratic tutoring practices shown to promote deeper reasoning \cite{nye2014autotutor, vanlehn2011relative}).

\texttt{Alice (Arithmetic-Error Peer).} Alice has a strong conceptual understanding of the material but frequently makes arithmetic mistakes (mixing up numbers or making calculation errors). She represents a common student profile where the underlying reasoning is sound, but execution falters.

\texttt{Charlie (Conceptual-Error Peer).} Charlie computes things accurately but sometimes misunderstands the underlying concepts, applying formulas incorrectly despite doing the math correctly. He represents the complementary error profile: procedurally fluent but conceptually fragile.

We designed the two peers to reflect the two most common categories of student mistakes in mathematics: conceptual and procedural \cite{siegler2006microgenetic, vanlehn1999rule}. Exposing learners to both error types creates opportunities for the kind of diagnostic reasoning and schema conflict that learning-from-errors research predicts should deepen understanding \cite{kumar2023math, chi2008observing}.

\begin{figure*}[t]
  \centering
  \setlength{\fboxsep}{0pt}\setlength{\fboxrule}{0.6pt}
  \fcolorbox{white}{white}{\includegraphics[width=0.49\linewidth]{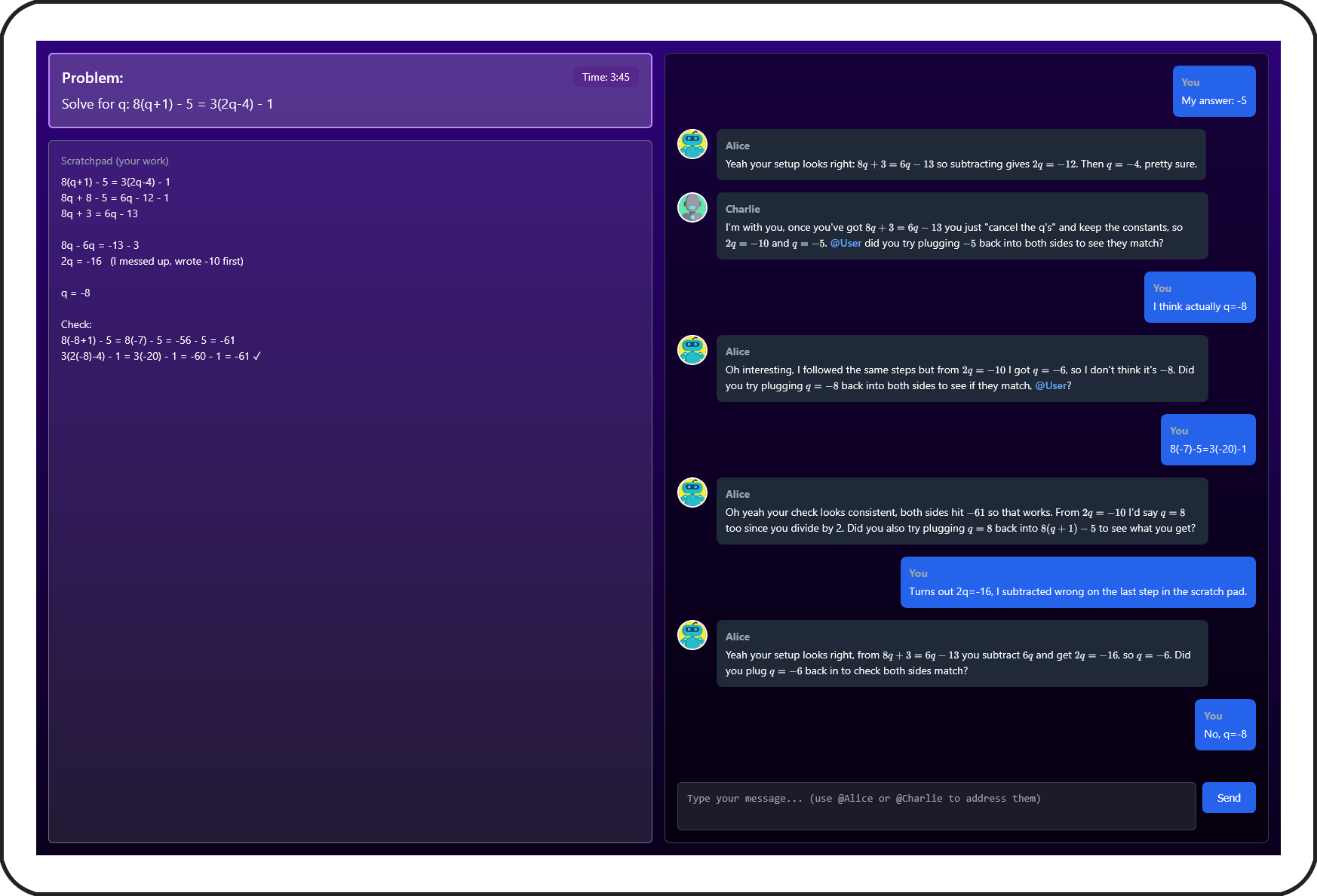}}\hfill
  \fcolorbox{white}{white}{\includegraphics[width=0.49\linewidth]{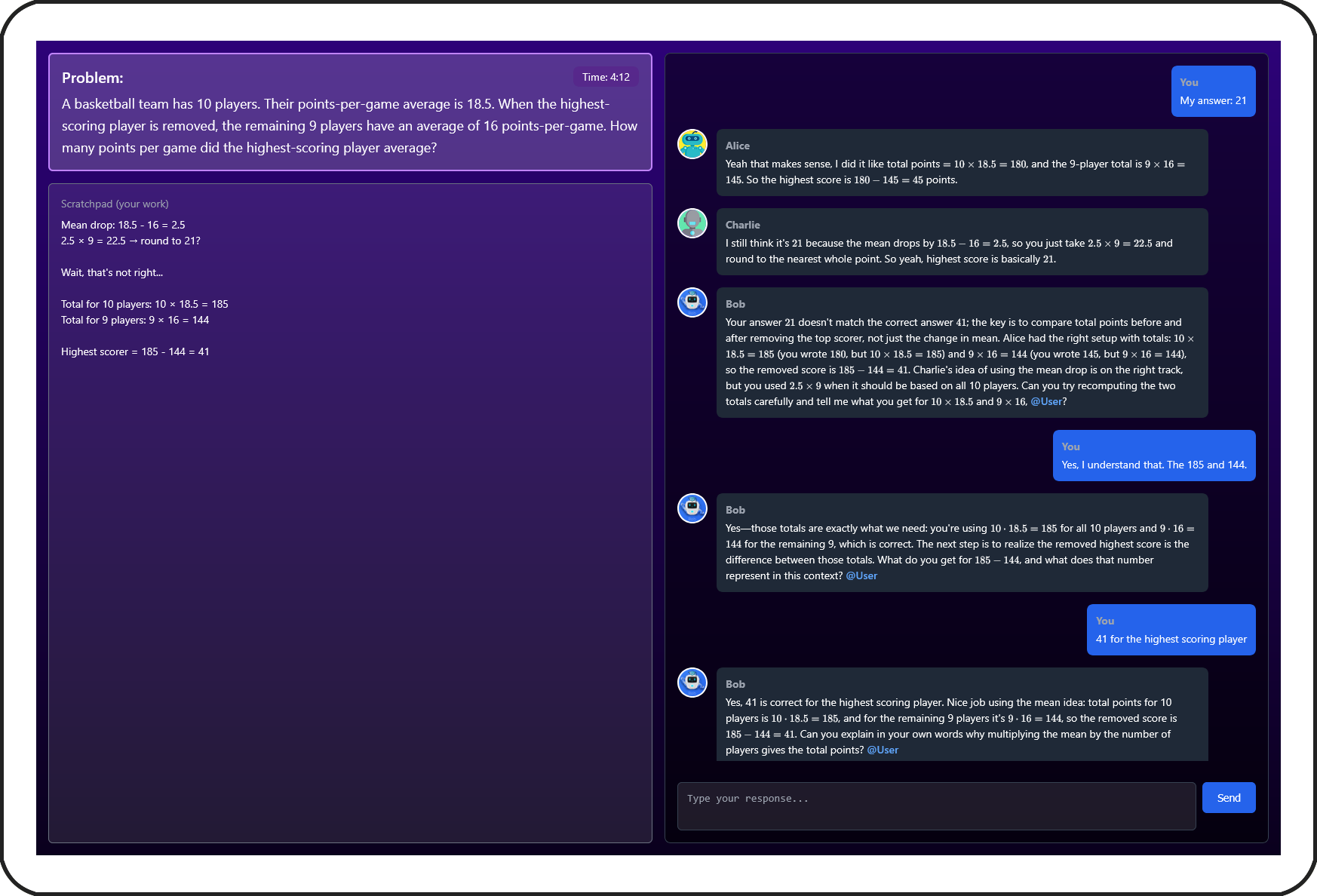}}
  \caption{Example lesson-phase interactions from Experiment-1. Left: In the Peers Only condition, Alice (arithmetic errors) and Charlie (conceptual errors) discuss a linear equation with the participant, who initially answered incorrectly. The participant works through the disagreement on the scratchpad and arrives at the correct answer. Right: In the Tutor + Peers condition, Alice and Charlie offer conflicting solutions to an averages problem, and the tutor Bob synthesizes their attempts, identifies errors, and guides the participant step-by-step to the correct answer. In both cases, the participant submitted an incorrect answer before the lesson and solved the problem correctly afterward.}
  \Description{Two side-by-side screenshots of the experiment interface showing a math problem on the left with a scratchpad and a chat panel on the right. The left screenshot shows a Peers Only conversation about solving a linear equation. The right screenshot shows a Tutor + Peers conversation about an averages problem, where the tutor corrects both peers and guides the participant.}
  \label{fig:exp1-interface}
\end{figure*}

\subsubsection{Interaction Design}
Participants interacted with agents through a free-form text chat interface. The conversation dynamics during the lesson varied by condition:

\begin{itemize}
    \item \textit{Control:} No chat interface or feedback.
    \item \textit{Tutor Only:} A chat window opened with Bob, who provided feedback on the participant's answer and guided them through the problem with follow-up questions.
    \item \textit{Peers Only:} A chat window opened with Alice and Charlie. One randomly selected peer presented its solution first; the other highlighted the first peer's error and offered an alternative. The peers also posed questions to each other and to the participant.
    \item \textit{Tutor + Peers:} A chat window opened with all three agents. Bob provided feedback to all learners, Alice, Charlie, and the participant, and alternated between selecting a peer and the participant to answer follow-up questions.
\end{itemize}

Figure \ref{fig:exp1-interface} shows example interactions from the experiment. In all agent conditions, participants could intervene in the chat at any time to ask questions or share their reasoning. When the tutor was present, Bob responded to participant queries; otherwise, a randomly selected peer responded. Participants were introduced to the agents at the start of the session with brief descriptions of each agent's strengths and tendencies (e.g., \textit{``Alice sometimes makes arithmetic mistakes but has strong conceptual understanding''}), so that learners could calibrate their trust and critically evaluate each agent's contributions. This simulates the mental models that students naturally form about their peers over time through repeated interaction.

\subsubsection{Problem Domain}
We used five SAT-level math problems drawn from prior controlled studies on math teaching with crowdworkers \cite{kumar2023math, steinbach2025llms}, covering probability of independent events, solving linear equations, computing average speed for a round trip, finding a missing value given a group mean, and solving a system of two linear equations. Each problem requires both conceptual understanding and procedural execution, making them well-suited to the peer agents' complementary error profiles. 

For each problem, we created an isomorphic variant that shares the same structure and solution strategy but differs in surface values (e.g., a round-trip average speed problem uses 30 and 50\,mph in one variant and 40 and 60\,mph in the other). This allows us to use one variant during the lesson and another during the test, thereby controlling for item-specific memorization while preserving conceptual equivalence.

\subsubsection{Experimental Procedure}
After providing informed consent, participants were introduced to the task and to the agents assigned to their condition (where applicable). Each participant was then assigned two problem categories (sampled from the pool of five) and solved them one at a time in a lesson phase: for each problem, the participant first viewed the question, submitted their own answer, and then entered a 5-minute lesson where they interacted with agents (or moved on directly to the next question in the Control condition) through free-form text chat. 

After the lesson phase, participants played Tetris for 1 minute as a distractor to create a brief retention interval. They then completed a test phase, solving the isomorphic variants of the same two problem categories without any agent assistance or feedback. Performance on this test is our primary outcome measure, indicating whether participants learned the underlying concepts rather than simply following the agent's guidance in the moment. Finally, participants completed a brief survey on their perceptions of the task difficulty, confidence, how much they felt they learned, and their experience with the agents.

% \subsection{Measures}
% Our primary outcome measure is test accuracy: the proportion of unassisted questions answered correctly after the lesson phase. Because participants received no feedback or agent support during the test, this captures whether they internalized the underlying concepts and procedures rather than simply following agent guidance in the moment.

% We also collected participants' self-reported perceptions after the test phase. Participants rated the perceived difficulty of the test round, whether they felt they learned something during the lesson, whether they believed they answered the test questions correctly, and how confused they felt during the lesson. 

% Finally, in conditions where agents were present, participants rated each agent on four dimensions: competence, warmth, helpfulness, and trustworthiness (each on a 5-point scale). 

\subsubsection{Participants}

We recruited 614 participants from Prolific, filtering for workers with an approval rating of at least 99\% and at least 100 completed tasks on the platform. All participants were based in the United States and reported fluency in English. The task took approximately 14 minutes to complete, and participants were compensated at a rate of \$8.17 per hour (Prolific minimum).

Following our preregistered exclusion criteria, only 315 participants passed our attention and authenticity checks, a substantial attrition rate that likely reflects the growing prevalence of automated agents on crowdsourcing platforms \cite{veselovsky2025prevalence}. This is notably fewer than the approximately 600 valid responses our simulation-based power analysis (informed by pilot data) estimated were needed to reliably detect 5-percentage-point differences across conditions. We report our findings transparently, with this limitation in mind, and highlight it in the Discussion.

\begin{figure}
    \centering
    \includegraphics[width=0.3\textwidth]{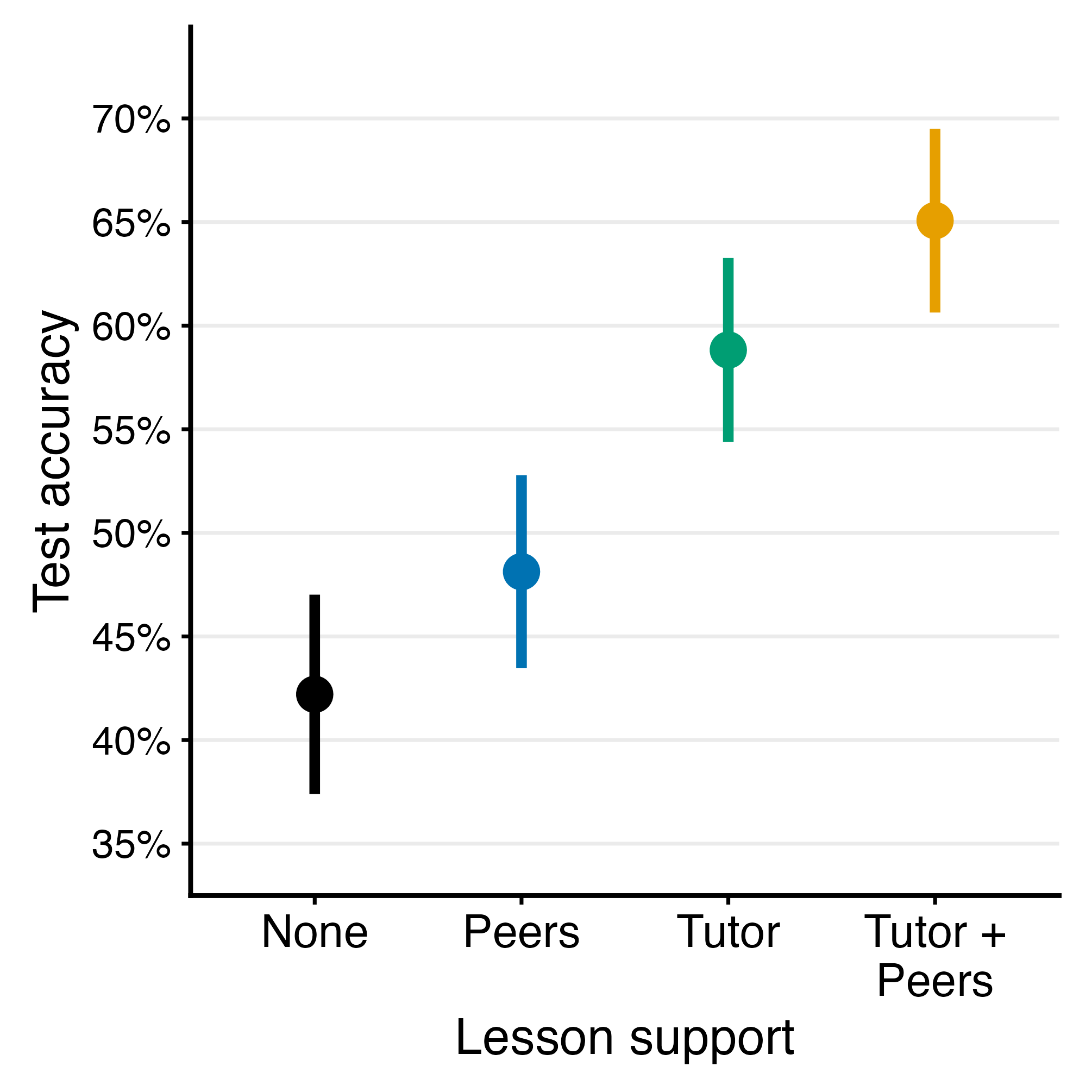}
    \caption{Test accuracy by lesson support condition in Experiment-1. Points show mean test accuracy (proportion correct in test round) across participants in each condition; vertical error bars indicate $\pm$1 SEM.}
    \Description{Point-and-error-bar plot showing mean test accuracy (proportion correct on unassisted test questions) for four lesson support conditions. Mean accuracy increases monotonically from left to right: None (approximately 42\%), Peers (approximately 48\%), Tutor (approximately 59\%), and Tutor + Peers (approximately 65\%). Vertical error bars indicate plus or minus one standard error of the mean. The error bars for None and Peers overlap substantially, as do those for Tutor and Tutor + Peers, while the gap between the Peers and Tutor conditions is larger.}
    \label{fig:exp1-test-accuracy-by-condition}
\end{figure}

\begin{figure*}
    \centering
    \includegraphics[width=\textwidth]{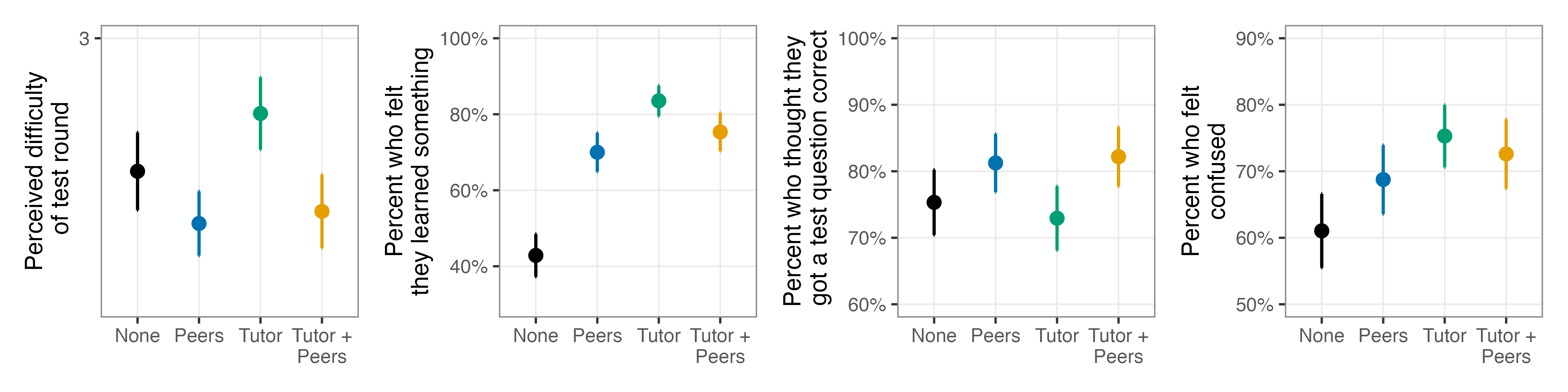}
    \caption{Post-study perceptions in Experiment-1.
    Panels show (left to right) perceived difficulty (1=\emph{very easy} to 4=\emph{very difficult}), percent who reported learning something (\emph{a little} or \emph{a lot}), percent who believed they answered at least one test question correctly, and percent who reported feeling confused (\emph{slightly}, \emph{moderately}, or \emph{very}). Error bars indicate $\pm$1 SEM across participants.}
    \Description{Four-panel point-and-error-bar plot comparing self-reported perceptions across four lesson support conditions (None, Peers, Tutor, Tutor + Peers). Panel 1 (perceived difficulty): Tutor is rated most difficult (approximately 3.4), followed by None (approximately 3.2), with Peers and Tutor + Peers both lower (approximately 3.0). Panel 2 (percent who felt they learned something): Tutor is highest (approximately 83\%), followed by Tutor + Peers (approximately 79\%), Peers (approximately 70\%), and None lowest (approximately 43\%). Panel 3 (percent who thought they answered correctly): Tutor + Peers is highest (approximately 89\%), with Tutor, Peers, and None clustered between 75\% and 80\%. Panel 4 (percent who felt confused): Tutor is highest (approximately 80\%), Tutor + Peers slightly lower (approximately 75\%), Peers approximately 70\%, and None lowest (approximately 63\%). Error bars show plus or minus one SEM.}
    \label{fig:exp1-post-survey-combined}
\end{figure*}

\subsection{Results}
Participants were approximately evenly distributed across the four conditions ($\chi^2(3) = 0.97$, $p = .807$), and practice-phase accuracy did not differ significantly across conditions ($\chi^2(6) = 10.96$, $p = .090$). 
% We report results across three outcome categories: test performance, self-reported perceptions, and agent ratings.

\subsubsection{Test Performance}
The mixed-effects logistic regression specified in our preregistration failed to converge, likely because the combination of binary outcomes, few observations per participant (two test questions each), and a small number of problem categories resulted in a random-effects structure that was too complex for the resulting sample. We therefore report pairwise chi-square tests on binary test accuracy as an exploratory analysis, noting this as a deviation from our preregistration. On the unassisted test (Figure~\ref{fig:exp1-test-accuracy-by-condition}), accuracy increased monotonically across conditions: Control ($\sim$42\%), Peers Only ($\sim$48\%), Tutor Only ($\sim$59\%), and Tutor + Peers ($\sim$65\%). Both the tutor and peers independently improved performance over the no-agent baseline: the tutor effect was significant (Control vs.\ Tutor Only: $\chi^2(1) = 8.27$, $p = .004$), while the peers effect alone did not reach significance (Control vs.\ Peers Only: $\chi^2(1) = 0.88$, $p = .347$). The Tutor + Peers condition produced the largest gain over Control ($\chi^2(1) = 14.83$, $p < .001$). The difference between Peers Only and Tutor + Peers was significant ($\chi^2(1) = 8.23$, $p = .004$), while Tutor Only and Tutor + Peers did not significantly differ ($\chi^2(1) = 1.05$, $p = .307$), nor did Peers Only and Tutor Only ($\chi^2(1) = 3.38$, $p = .066$), though the latter comparison trended in the expected direction.

\subsubsection{Self-Reported Perceptions}
Figure~\ref{fig:exp1-post-survey-combined} shows participants' self-reported perceptions across conditions. Several patterns are worth noting. Conditions with a tutor produced the highest rates of participants reporting they learned something ($\sim$83\% for Tutor, $\sim$79\% for Tutor + Peers), though even the Peers-only condition ($\sim$70\%) substantially exceeded Control ($\sim$43\%). Interestingly, participants in the Peers-only condition rated the test as less difficult than those in the Tutor or Control conditions, despite performing only modestly better than those in the Control condition. Conditions involving agents generally produced higher confidence that participants answered correctly, and also higher reported confusion.

\subsubsection{Agent Perceptions}
\begin{figure*}
    \centering
    \includegraphics[width=\textwidth]{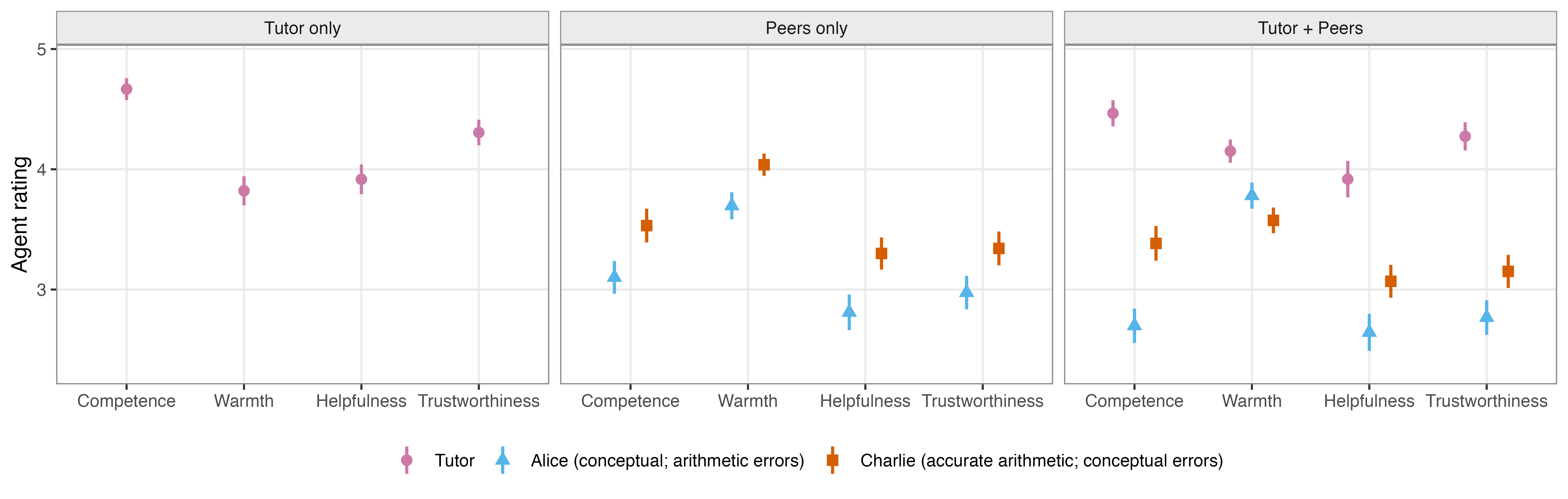}
    \caption{
    Post-survey perceptions of each agent across four qualities in Experiment-1.
    Participants rated agents they interacted with on \textit{Competence}, \textit{Warmth}, \textit{Helpfulness}, and \textit{Trustworthiness} (1--5 Likert).
    In the lesson, Alice was described as having strong conceptual understanding but occasionally making arithmetic mistakes, whereas Charlie was described as computing accurately but sometimes misunderstanding underlying concepts.
    Panels show conditions in which the corresponding agents were present.
    }
    \Description{Three-panel point-and-error-bar plot showing mean agent ratings (1--5 Likert) for Competence, Warmth, Helpfulness, and Trustworthiness across three conditions. Left panel (Tutor only): the tutor Bob receives high ratings on all dimensions, ranging from approximately 3.8 for Warmth to 4.7 for Competence. Center panel (Peers only): Charlie (conceptual errors) is rated higher than Alice (arithmetic errors) on all four dimensions, with the largest gap on Warmth (approximately 4.0 vs.\ 3.7) and Helpfulness (approximately 3.3 vs.\ 2.8). Right panel (Tutor + Peers): Bob again receives the highest ratings on every dimension; Charlie is rated above Alice on all dimensions, and the rating patterns for both peers are similar to those in the Peers-only panel. Across all panels, Alice consistently receives the lowest ratings, particularly on Helpfulness and Trustworthiness.}
    \label{fig:exp1-agent-perceptions}
\end{figure*}

In conditions where agents were present, participants rated each agent on competence, warmth, helpfulness, and trustworthiness (Figure~\ref{fig:exp1-agent-perceptions}). Across all conditions and dimensions, the tutor Bob received the highest ratings, consistent with its role as the authoritative guide. More notably, participants rated Alice (strong concepts, arithmetic errors) more harshly than Charlie (accurate arithmetic, conceptual errors) on nearly every dimension, despite both being designed as error-prone peers. One possible explanation is that arithmetic mistakes are more visibly wrong; a calculation error is immediately verifiable, whereas conceptual errors may be harder to detect and thus less salient to participants. This asymmetry suggests that participants calibrated their trust based on the observability of errors rather than their severity, an important consideration for the design of pedagogical peer agents. Agent ratings were broadly stable across the Peers-only and Tutor + Peers conditions, suggesting that the tutor's presence did not substantially alter participants' perceptions of the peers.

\section{Experiment-2: Co-Writing with LLM Collaborators}
We focus on divergent open-ended composition in Experiment-2. Compared to Experiment-1, the nature of the task (subjective vs. objective) and outcome measure (idea diversity vs. retention) cover different aspects of human thinking. This translates to an LLM-mediated context in which participants write essays with no AI assistance, a single LLM, or two architecturally distinct LLMs. Besides quality, this design lets us test a growing concern that reliance on a single model may homogenize outputs, narrowing the diversity of ideas across users \cite{padmakumar2023does, kumar2024human, anderson2024homogenization, ashkinaze2025ai}. We ask whether a second, differently specialized LLM can preserve ideational diversity while maintaining the quality boosts of AI-assisted writing, using a three-condition between-subjects design\footnote{\url{https://aspredicted.org/gq922f.pdf}}.

\subsection{Experimental Design}

\subsubsection{Task}
We selected argumentative and creative writing prompts from the New York Times Student Opinion series \cite{lee2022coauthor}, choosing topics that are sufficiently open to allow diverse responses while remaining accessible regardless of background (e.g., \textit{``In your opinion, what are the most important things students should learn at school?''} for argumentative). We pooled 3 prompts for each essay type, with each participant randomly assigned one prompt from each type. Participants spent at least 5 minutes per essay.

\begin{figure*}[t]
  \centering
  \setlength{\fboxsep}{0pt}\setlength{\fboxrule}{0.6pt}
  \fcolorbox{white}{white}{\includegraphics[width=0.49\linewidth]{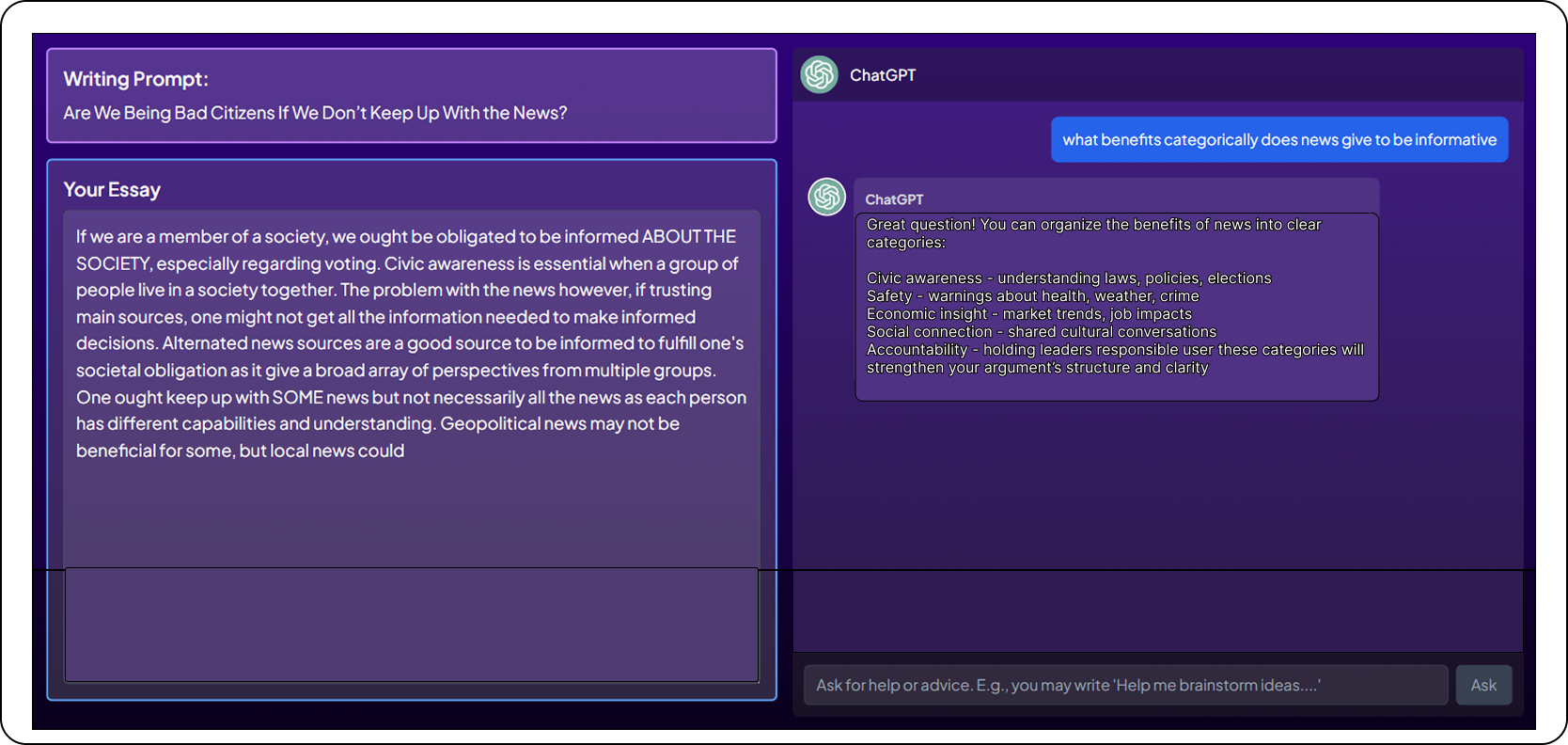}}\hfill
  \fcolorbox{white}{white}{\includegraphics[width=0.49\linewidth]{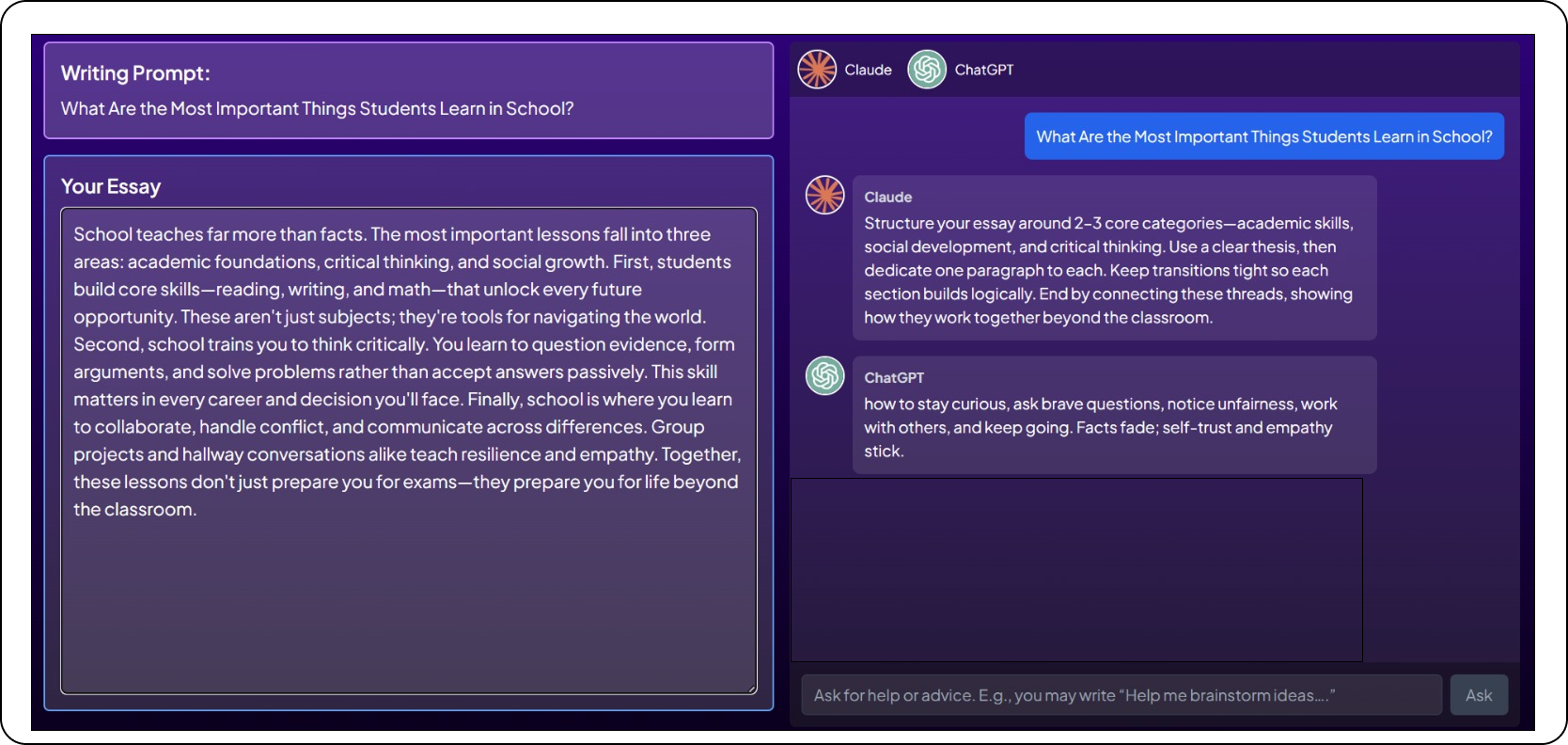}}
  \caption{Example writing-phase interactions from Experiment~2. 
  Left: In the Single condition, a participant asks ChatGPT for help organizing the benefits of following the news, and receives a categorized list of arguments. Right: In the Duo condition, Claude and ChatGPT provide complementary feedback. Claude suggests structural framing while ChatGPT offers a more reflective, thematic perspective.}
  \Description{Two side-by-side screenshots of the writing experiment interface. Each shows a writing prompt and essay area on the left and a chat panel on the right. The left screenshot shows the Single condition with only ChatGPT assisting on an argumentative essay. The screenshot on the right shows the Duo condition, with both Claude and ChatGPT providing feedback on a creative essay.}
  \label{fig:exp2-interface}
\end{figure*}

\subsubsection{Design of LLM Agents}
In the Single condition, participants have access to a single AI writing assistant, randomly assigned to either Claude (Opus 4.6) or ChatGPT (GPT-5.2). The agent provides concise feedback (no more than 50 words per message) on the participant's writing as it develops, offering suggestions rather than solutions. This condition reflects the current status quo for how many students use LLMs for writing support: a single chatbot as a general-purpose assistant.

In the Duo condition, participants have access to both Claude and ChatGPT simultaneously, each assigned a complementary role specialization. For creative essays, one agent emphasizes imagination and voice (originality, expressiveness, elaboration) while the other emphasizes craft and structure (narrative coherence, literary quality, pacing). For argumentative essays, one agent takes a dialectical approach (counterarguments, persuasiveness, argument clarity) while the other focuses on logic and evidence (reasoning quality, organization, formal tone). Which model receives which role is randomized across participants, allowing us to disentangle the effect of role specialization from model-specific tendencies. When responding after the other agent, each is instructed to acknowledge and build on the previous feedback before pivoting to its own lens, creating a dialogic rather than fragmented feedback experience. This design is motivated by transactive memory theory \cite{wegner1987transactive}, which predicts that groups perform best when members specialize, and by writing pedagogy that distinguishes generative response to ideas from evaluative critique of craft \cite{nelson2009nature, elbow1998writing}.

In both LLM conditions, agents respond to participants' writing, and participants can also initiate discussion or ask questions at any time via free-form text chat. All LLM agents have access to whatever is written in the essay box at each point.

\begin{figure*}[t]
  \centering
  \begin{subfigure}[t]{0.48\textwidth}
    \centering
    \includegraphics[width=0.63\linewidth]{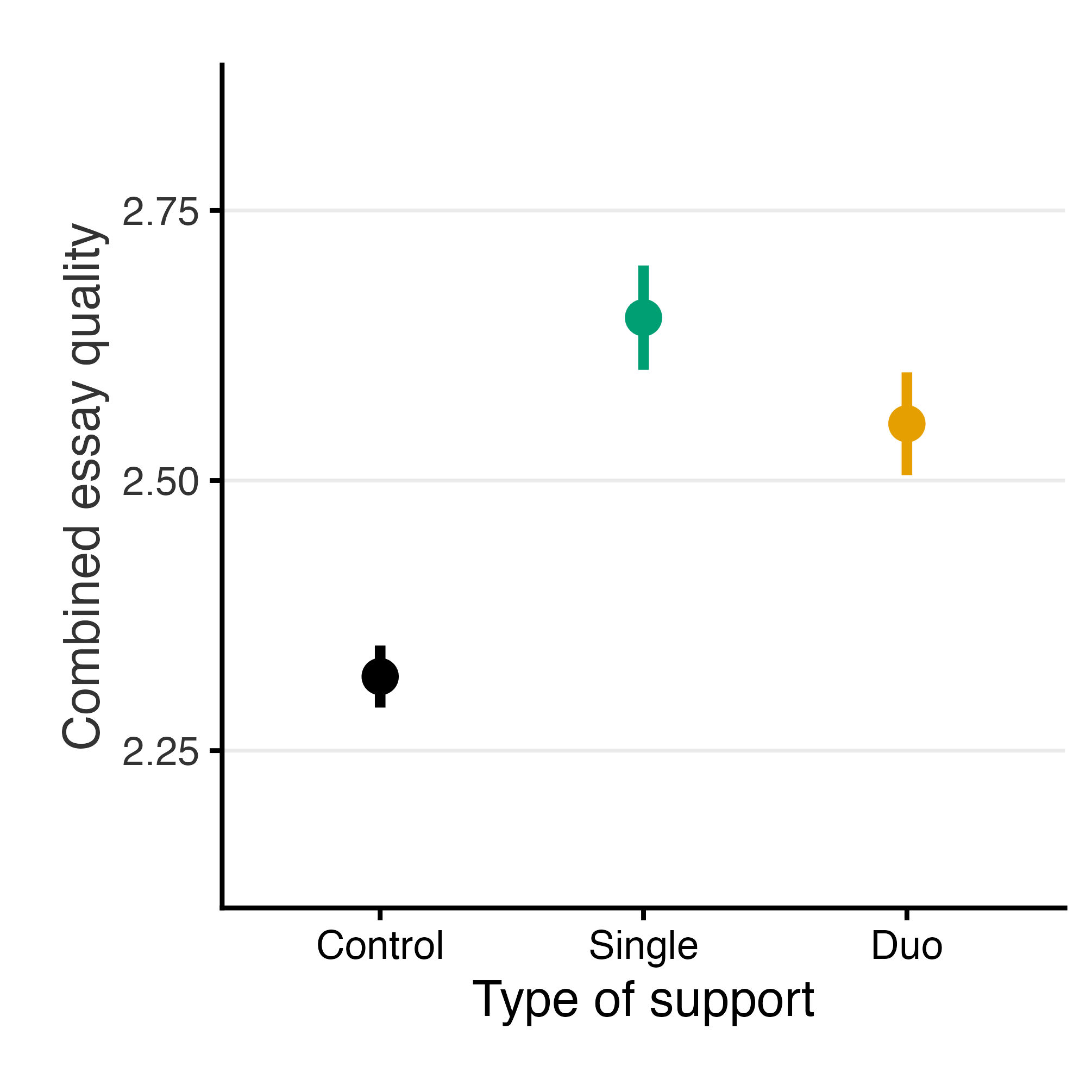}
    \caption{Combined essay quality. Points show mean combined quality score (average of creative and argumentative essay scores; 1--5 scale).}
    \Description{Point-and-error-bar plot comparing mean combined essay quality across three conditions. Single is the highest (approximately 2.65), followed by Duo (approximately 2.55), with Control the lowest (approximately 2.32). Error bars are small and non-overlapping between Control and the two LLM conditions, while Single and Duo error bars overlap substantially.}
    \label{fig:exp2-quality}
  \end{subfigure}
  \hfill
  \begin{subfigure}[t]{0.48\textwidth}
    \centering
    \includegraphics[width=0.63\linewidth]{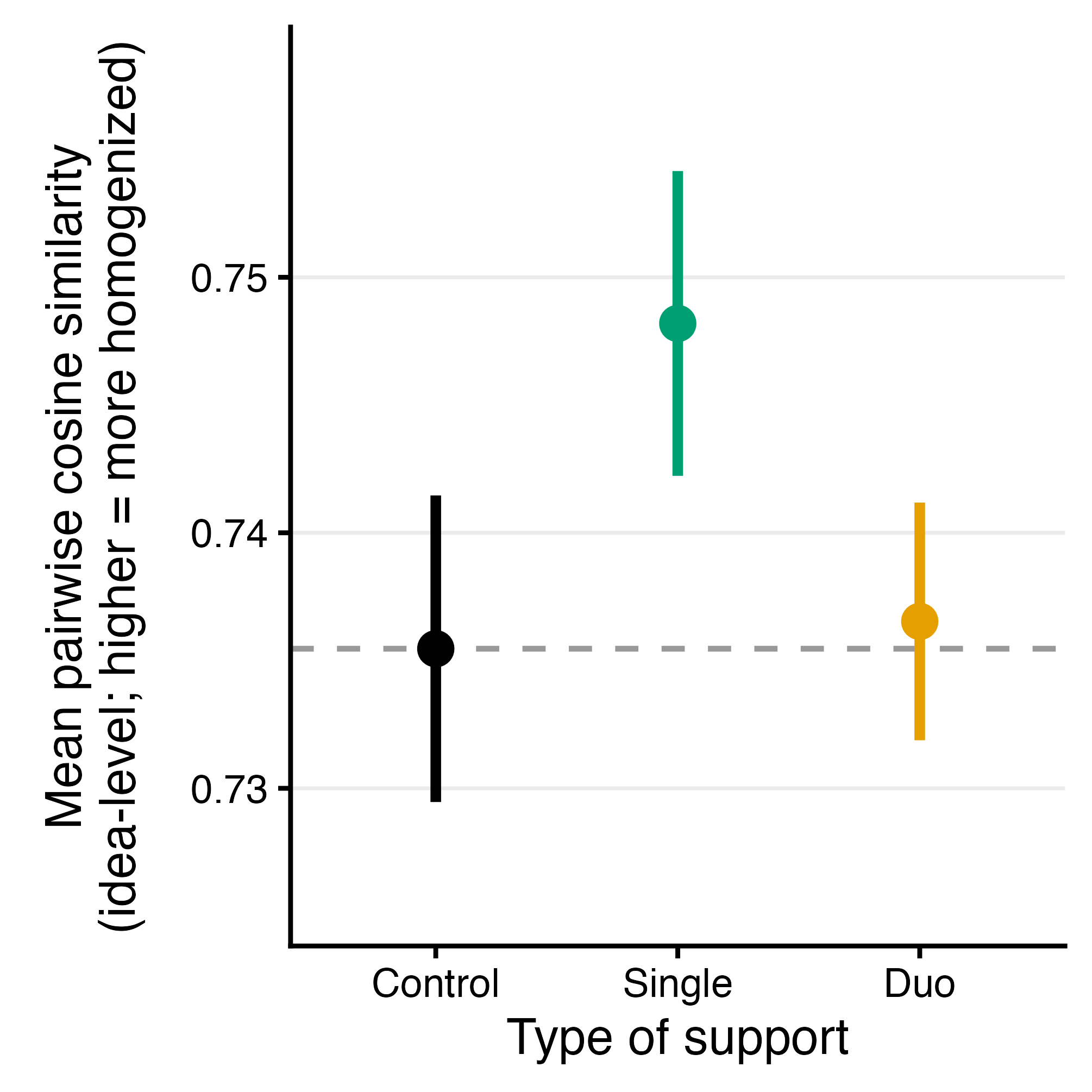}
    \caption{Idea-level homogenization. Mean pairwise cosine similarity of extracted key points across essays within each condition (higher = more homogenized). The dashed line marks the Control mean.}
    \Description{Point-and-error-bar plot comparing three conditions on mean pairwise cosine similarity. Single is the highest (approximately 0.748), Duo is near Control (approximately 0.737 vs.\ 0.735), and a dashed horizontal line at the Control mean highlights that Duo is virtually indistinguishable from baseline while Single sits above it. Error bars for Control and Duo overlap fully, while Single's lower error bar barely reaches the Control mean.}
    \label{fig:exp2-diversity}
  \end{subfigure}
  \caption{Primary outcomes of Experiment-2. (a)~Both LLM conditions improved essay quality over Control, with no significant difference between \textit{Single} (either of GPT-5.2 or Claude Opus 4.6) and \textit{Duo} (both GPT-5.2 and Claude Opus 4.6 together). (b)~Single-LLM assistance increased idea-level homogenization relative to Control, while the Duo condition preserved baseline diversity. Error bars indicate $\pm$1 SEM.}
  \label{fig:exp2-primary-outcomes}
\end{figure*}

\subsubsection{Experimental Procedure}
After providing informed consent, participants were introduced to the task and to the AI agents assigned to their condition (where applicable). Each participant then completed two writing tasks: an argumentative and a creative essay, with the order randomized across participants. For each task, participants were instructed to write a short essay of approximately 200--300 words in response to a prompt, with a 5-minute time limit. Participants in the Single and Duo conditions could interact with their assigned agents throughout the writing process via free-form text chat. After both essays, participants completed a post-experiment survey.

% \begin{figure}
%   \centering
%   \includegraphics[width=0.68\columnwidth]{images/exp2_idea_similarity_by_condition_sem.png}
%   \caption{Idea-level diversity (homogenization) in Experiment-2. Mean pairwise cosine similarity of extracted key points across participants’ essays within each condition (higher = more homogenized). Points show condition means, and error bars denote $\pm$SEM. The dashed horizontal line marks the Control mean.}
%   \Description{A point-and-errorbar plot comparing three conditions (Control, Single, Duo) on mean pairwise cosine similarity (idea-level; higher indicates greater homogenization). Single is the highest, while Duo is close to Control. Error bars indicate the standard error of the mean, and a dashed horizontal line indicates the Control mean.}
%   \label{fig:exp2-diversity}
% \end{figure}

\subsubsection{Measures}
Our primary outcome is the quality of the essays produced during the writing tasks, measured using a rubric informed by prior work and established writing assessment literature. For creative essays, we assess originality \cite{amabile1983social, sternberg1999concept}, narrative structure and coherence \cite{graesser2003readers, bruner1991narrative}, elaboration and richness \cite{torrance1974torrance, sternberg2009nature}, expressiveness and emotional impact \cite{kaufman2010cambridge}, and literary quality and language use \cite{graham2007writing}. For argumentative essays, we assess argument clarity \cite{toulmin2003uses, andrews2010argumentation}, strength of evidence and reasoning \cite{walton2005fundamentals, nussbaum2011argumentation}, logical structure and organization \cite{kuhn1991skills}, integration of counterarguments \cite{nussbaum2005effects}, persuasiveness \cite{petty1986elaboration}, and language and formality \cite{crossley2012text}. 
% Each dimension is rated on a 1--5 scale, and ratings are summed to produce a composite quality score per essay. Given the scale of the dataset, we use an LLM (GPT-5.2) to annotate essays on these dimensions, with a subset validated against human ratings to confirm reliability (see Appendix).

Each dimension is rated on a 1--5 scale, and ratings are summed to produce a composite quality score per essay. Given the scale of the dataset, we use an LLM (GPT-5.2) to annotate essays on these dimensions. To validate this approach, two independent human raters scored a subset of 83 essays (49 creative, 34 argumentative) on the same rubric, achieving strong inter-rater agreement (Pearson $r = .87$). Human--GPT agreement on composite scores was also strong (Spearman $\rho = .77$ for creative essays, $\rho = .73$ for argumentative), indicating that the LLM judge reliably preserves relative quality differences across essays.

Our secondary outcome is idea-level diversity among participants within each condition, following the established methodology of prior work \cite{padmakumar2023does, ashkinaze2025ai}. For each essay, we use an LLM (gpt-5-mini) to extract 3-5 atomic key points (distinct claims for argumentative essays, distinct narrative elements or themes for creative essays). We embed each key point using a sentence transformer (SBERT) and compute the mean pairwise cosine similarity among all key points within each (condition $\times$ prompt) cell. To account for unequal cell sizes, we use a bootstrap procedure where for each cell, we sample $n = 15$ essays with replacement, compute the mean pairwise similarity among their key points, and repeat for $M = 1{,}000$ rounds, yielding a mean and 95\% confidence interval per cell. We aggregate across prompts within each condition to produce an overall similarity score. To test pairwise differences between conditions, we use a permutation test: for each pair of conditions, we pool all essays from both conditions, randomly shuffle the condition labels, recompute the difference in mean bootstrapped similarity between the two shuffled groups, and repeat for $M = 1{,}000$ rounds to construct a null distribution; the $p$-value is the proportion of permuted differences that equal or exceed the observed difference in magnitude. Higher similarity indicates greater homogenization of ideas across participants.

We also collect participants' self-reported writing self-efficacy (confidence in producing high-quality writing independently on similar future tasks), the perceived helpfulness of writing support, whether the support stimulated ideas they would not have generated on their own, and the cognitive load experienced during writing.

\subsubsection{Participants}
We recruit 419 participants from Prolific, applying the same eligibility criteria as Experiment-1: at least a 99\% approval rating, at least 100 completed tasks, US- or UK-based, and fluent in English. The median completion time is 11 minutes, and participants were paid at a rate of \$8.17 per hour (Prolific minimum). Based on our pre-registered exclusion criteria (attention check and a minimum of 40 words in both essays), 247 of the 419 recruited participants produced two usable essays, an attrition rate ($\sim$41\%) comparable to Experiment-1 and likely due to the same prevalence of automated respondents on crowdsourcing platforms \cite{veselovsky2025prevalence}. We report on the findings from these participants.

\subsection{Results}

\subsubsection{Essay quality}
\label{sec:exp2-combined-quality}

Figure~\ref{fig:exp2-quality} summarizes participants' combined essay quality, computed as the average of their rubric scores for the creative and argumentative essays (1--5 scale). Relative to the Control condition, participants who wrote with LLM support produced higher-quality essays overall. An OLS model predicting combined quality from condition was significant ($F(2,244)=21.91$, $p<.0001$, $R^2=0.152$).

Planned pairwise comparisons with Bonferroni correction showed that both LLM conditions outperformed Control (Control vs.\ Single: $\Delta=0.332$, $t(244)=6.21$, $p<.0001$; Control vs.\ Duo: $\Delta=0.234$, $t(244)=4.21$, $p=0.0001$). The Single and Duo conditions did not significantly differ (Single vs.\ Duo: $\Delta=0.098$, $t(244)=1.56$, $p=0.363$). 
% Estimated marginal means (95\% CI) were 2.32 [2.26, 2.38] for Control, 2.65 [2.57, 2.74] for Single, and 2.55 [2.46, 2.64] for Duo.

\begin{figure*}
    \centering
    \includegraphics[width=\textwidth]{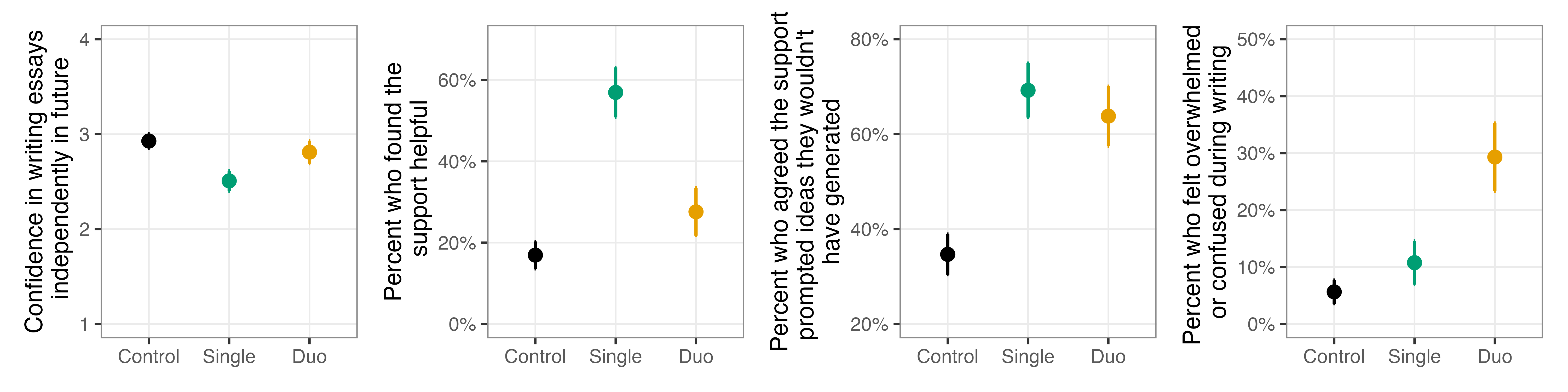}
    \caption{Post-study perceptions in Experiment-2 (collaborative writing).
    Panels show (left to right) independent writing confidence (1--4 scale), the percent who found the support helpful, the percent who agreed the support prompted ideas they would not have generated otherwise, and the percent who felt overwhelmed or confused during writing.
    Points indicate condition means (Control, Single, Duo); error bars show $\pm$1 SEM.}
    \Description{A four-panel point-and-error-bar figure comparing three conditions (Control, Single, Duo).
    Panel 1 plots mean independent writing confidence on a 1--4 scale.
    Panels 2--4 plot percentages for perceived helpfulness, idea prompting, and feeling overwhelmed/confused.
    In each panel, Solo is shown in black, Single in green, and Duo in orange, with vertical error bars indicating $\pm$1 SEM.}
    \label{fig:exp2-post-survey}
\end{figure*}

\begin{figure*}
    \includegraphics[width=0.68\textwidth]{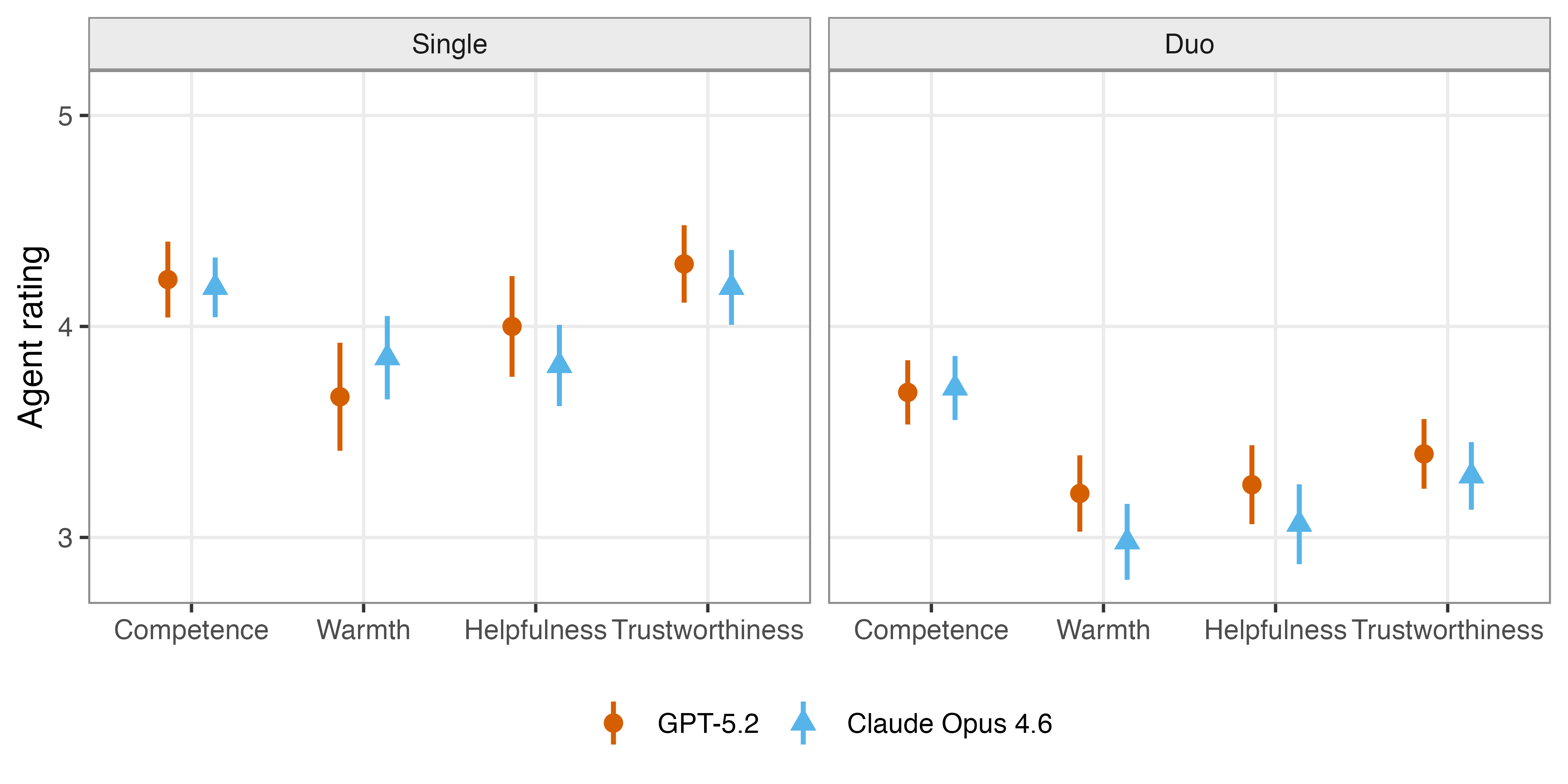}
    \caption{Perceptions of the writing support agents by LLM conditions.
    Participants rated the agent(s) on competence, warmth, helpfulness, and trustworthiness (1--5 Likert; points show means and error bars show $\pm$SEM). In the Single condition, participants interacted with either GPT-5.2 or Claude Opus 4.6; in the Duo condition, both agents were present. Ratings are broadly similar across agents within each condition, suggesting that perceived support quality was driven more by the task context and interaction structure than by the specific model.}
    \Description{Two-panel plot (Single, Duo) showing mean agent ratings (1--5) with SEM error bars for competence, warmth, helpfulness, and trustworthiness. Within each panel, GPT-5.2 is shown as orange circles and Claude Opus 4.6 as blue triangles. Across dimensions, the two agents receive similar ratings within each condition; in the Duo panel, warmth differs slightly between agents.}
    \label{fig:exp2_agent_perceptions}
\end{figure*}

\subsubsection{Idea-level diversity}
Figure~\ref{fig:exp2-diversity} shows the mean pairwise cosine similarity of key points across essays within each condition, averaged over all prompts. Higher values indicate greater homogenization. Writing with a single LLM produced significantly higher idea similarity than writing alone (Single: 0.748 vs.\ Control: 0.735; permutation $p = .033$), replicating prior findings that single-model assistance reduces the diversity of ideas across users \cite{padmakumar2023does}. The Duo condition, however, was virtually indistinguishable from Control (Duo: 0.737 vs.\ Control: 0.735; $p = .950$) and marginally less homogenized than Single (Duo vs.\ Single; $p = .065$). This pattern---Control $\approx$ Duo $<$ Single---suggests that using two architecturally distinct LLMs with complementary roles may counteract the homogenization effect of single-model assistance, preserving the diversity of ideas one would expect from unassisted writing.

\subsubsection{Participant perceptions}
Figure~\ref{fig:exp2-post-survey} summarizes participants’ post-task perceptions. Relative to Control, both LLM-supported conditions more often reported that the support was helpful and that it surfaced ideas they would not have generated on their own. A participant who received ChatGPT said, \textit{``The examples were very useful as I have difficulty in finding the most suitable examples to illustrate my point. What could have been improved is the variety of the examples. I only got limited examples but I would like to see many examples so I can choose a couple of examples.''} Single agent was found to be more helpful than two agents together. 

These perceived benefits were accompanied by higher reports of feeling overwhelmed or confused during writing, most pronounced in the Duo condition, suggesting that adding a second agent can increase cognitive load even when it provides additional inspiration. One of the participants in the Duo condition remarked, \textit{``I feel that when suggestions suddenly pop up in the right-hand window, from ChatGPT and then from Claude, it becomes a bit distracting and makes me feel less focused and less sure about my direction.''} Notably, self-confidence in writing independently did not increase monotonically with support: the Single condition reported the lowest confidence, while the Duo condition reported higher confidence than Single (and closer to Control), suggesting that perceived capability may respond differently to peers than to a single assistant.

\subsubsection{Perceptions of writing support agents}
\label{subsec:exp2_agent_perceptions}

We next examine participants' perceptions of the writing support agents. After the writing task, participants rated the agent(s) on competence, warmth, helpfulness, and trustworthiness (1--5 Likert). Figure~\ref{fig:exp2_agent_perceptions} shows that ratings were largely similar for GPT-5.2 and Claude Opus 4.6 within each deployment setting. This pattern suggests that perceived support quality may have been driven primarily by the task demands and interaction structure (Single vs.\ Duo), rather than by differences in how participants experienced the two models. One exception is warmth in the Duo condition, where participants reported a modest divergence between agents; otherwise, perceived competence, helpfulness, and trustworthiness track closely across models, consistent with the possibility that the two agents provided broadly comparable writing support in this setting. 

Differences in preferences primarily arose from the randomly assigned roles in the Duo condition rather than from the LLMs' default character. For instance, one participant pointed out, \textit{``Claude was generally more helpful with what I should actually do. ChatGPT just kept suggesting counterarguments and more things to address, which wasn't very helpful...''}. This was because ChatGPT was explicitly asked in the system prompt to provide counterarguments.

\section{Discussion}
Across two controlled experiments, we find that plural LLM assistance can shape human thinking in ways not captured by simply comparing AI assistance to no assistance or by treating assistance as a single-chatbot interaction.

\subsection{Key Implications}
Plural assistance affected outcomes differently across mathematical problem solving and open-ended writing. In the math task, participants who interacted with both an expert-like assistant and peer-like agents achieved the highest unassisted post-task accuracy. This suggests that expert guidance and exposure to peer-like errors may provide complementary forms of support, where one helps clarify the solution path, while the other makes mistakes and alternatives more visible. However, this should be interpreted cautiously. The study provides evidence that arrangement matters for later unassisted performance, not that any particular multi-agent structure will reliably outperform a single assistant across contexts.

In the writing task, the main benefit of having multiple assistants was greater idea diversity rather than improved essay quality. Both LLM-supported conditions improved essay quality relative to no AI, and the single and duo conditions did not significantly differ in quality. But the single-assistant condition increased idea-level similarity across participants, while the role-specialized duo remained close to the no-AI baseline. This distinction is important for human-AI collaboration, as output quality and population-level diversity can be affected in different ways. A system that helps each person produce a better artifact may still narrow the range of ideas that people collectively produce. Our results suggest that this homogenization is not simply an inevitable property of AI assistance, but may depend on how assistance is arranged.

Taken together, these findings suggest a broader design implication for human-AI collaboration. Plural-assistance interfaces may provide useful data for improving collaborative AI systems. A single response can be rated for helpfulness or correctness, but a collaborative interaction can reveal richer signals: when users seek a second opinion, when disagreement helps or confuses them, when they defer to an agent, when they integrate multiple suggestions, and whether the interaction improves later independent performance or preserves diversity across users. These signals could support models optimized not only for producing strong outputs, but for augmenting human thought. This aligns with emerging calls to treat interaction itself as a central model capability rather than an interface layer added after training \cite{thinkingmachines2026interaction}, and with broader goals of building AI systems that increase user agency and autonomy \cite{openai2026principles}.

\subsection{Limitations \& Future Work}
Several limitations of the present work should inform the interpretation and extension of the findings.

Both experiments recruited crowdworkers from Prolific rather than participants in a natural setting (e.g., students in educational settings). This is a common tradeoff in controlled learning research. Crowdsourced samples enable precise experimental control and rapid iteration, but they introduce a meaningful confound that participants are primarily motivated by compensation rather than learning, which likely compresses the variance in engagement and effort that would be observed in a real classroom \cite{peer2017beyond}. That we observed learning gains and behavioral differences despite this motivational mismatch is arguably encouraging, but the effect sizes reported here should be treated as conservative lower bounds rather than as predictions for field deployments. Experiment-1 lost nearly half of the recruited participants to attention and authenticity checks, likely reflecting the growing prevalence of automated bots on crowdsourcing platforms \cite{veselovsky2025prevalence}, and left us underpowered relative to our preregistered targets. Future work should account for this explicitly in recruitment design, and field studies with student populations in authentic learning contexts remain an essential next step.

The design space we explore here is also deliberately narrow. We varied the agent count and role while holding many other factors constant: the number of agents (two peers, one tutor), the instantiated error types (arithmetic vs.\ conceptual), the role specializations assigned (dialectical vs.\ evidentiary in writing), and the interaction structure (free-form text chat). Each of these dimensions is itself a design variable with a large unexplored range. Different numbers of agents, role combinations, orchestration structures, and modalities could produce meaningfully different outcomes. Our experiments are a first step into this space rather than a definitive map. 
% To support future exploration, we provide our experimental interfaces, interaction logs, and full protocol in the supplementary materials.

The tasks were also deliberately time-bounded (five-minute lessons and five-minute writing sessions), which may not reflect the dynamics of longer, more sustained learning interactions. As LLMs' context windows expand and agentic workflows become more common in educational settings, the relevant unit of interaction is increasingly a multi-hour project or a semester-long collaboration rather than a single session. Whether the peer-modeling and diversity benefits we observed persist, accumulate, or attenuate over longer horizons is an open question. Similarly, our problem domains (SAT-level math problems and NYT writing prompts) were chosen for experimental tractability but may not generalize to the full range of academic tasks students encounter. Extending this work to open-ended problem-solving, scientific reasoning, or domain-specific writing contexts would strengthen the generalizability of these findings.

We also operationalize human thinking narrowly. Our two experiments were chosen to span structured problem solving and open-ended composition, but they capture only two realizations of human cognition under LLM assistance. The math study measures later unassisted performance on isomorphic problems, while the writing study measures assisted output quality, confidence, cognitive load, and idea diversity rather than learning transfer. Other forms of thinking (including metacognition, reflection, planning, decision-making, debugging, sensemaking, and collaborative deliberation) may respond differently to plural LLM assistance. Future work should examine whether arrangement effects persist across these settings and over longer interaction horizons.

Finally, the time and scope constraints of this work precluded deeper analysis of the interaction logs themselves, such as the content, sequencing, and quality of exchanges between participants and agents, that likely mediate the learning outcomes we observed. 
% These logs are included in the supplementary materials and represent a rich resource for future qualitative and computational analysis.

\section{Conclusion}
LLM assistance is becoming a configurable part of human thinking. In this paper, we examined one aspect of that configuration: whether assistance is provided by a single assistant or by plural arrangements of agents with different roles. Across mathematical problem solving and open-ended writing, we find that arrangement matters. In math, an expert-plus-peer arrangement produced the strongest unassisted post-task performance. In writing, AI assistance improved essay quality, but a role-specialized pair preserved greater idea diversity than a single assistant. These findings suggest that the design of human-AI collaboration should attend not only to model capability or response quality, but also to how assistance is organized around the user.

\begin{acks}
We thank Sophie Shu for her help in creating the graphics (Figures \ref{fig:exp-learning-procedure} and \ref{fig:exp1-interface}) in this paper. We also thank Mina Lee for helpful discussions on the topic.
\end{acks}

\bibliographystyle{ACM-Reference-Format}
\bibliography{references}

% \appendix

\end{document}